\documentclass[%
 aip,
 amsmath,amssymb,
 reprint,%
]{revtex4-1}

\usepackage{graphicx}% Include figure files
\usepackage{dcolumn}% Align table columns on decimal point
\usepackage{bm}% bold math
\usepackage[utf8]{inputenc}
\usepackage[T1]{fontenc}
\usepackage{mathptmx}

\usepackage{natbib}
\usepackage{textcomp}
\usepackage{subfigure}
\usepackage{mwe}
\usepackage{color}
\usepackage[normalem]{ulem}
\usepackage{soul}
\usepackage{cancel}

\begin{document}

\preprint{AIP/123-QED}

\title{X-ray photons  produced from a plasma-cathode electron beam for radiation biology applications}
% Force line breaks with \\

\author{F.Gobet}
 \email{gobet@cenbg.in2p3.fr}
\affiliation{Universit\'e Bordeaux, CNRS-IN2P3, CENBG, F-33175 Gradignan, France}
\author{P.Barberet}
\affiliation{Universit\'e Bordeaux, CNRS-IN2P3, CENBG, F-33175 Gradignan, France}
\author{L.Courtois}%
\affiliation{CEA, CESTA, F-33116 Le Barp, France}
\author{G.Deves}
\affiliation{Universit\'e Bordeaux, CNRS-IN2P3, CENBG, F-33175 Gradignan, France}
\author{J.Gardelle}%
\affiliation{CEA, CESTA, F-33116 Le Barp, France}
\author{S.Leblanc}%
\affiliation{Universit\'e Bordeaux, CNRS-IN2P3, CENBG, F-33175 Gradignan, France}
\author{L.Plawinski}
\affiliation{Universit\'e Bordeaux, CNRS-IN2P3, CENBG, F-33175 Gradignan, France}
\author{H.Seznec}
\affiliation{Universit\'e Bordeaux, CNRS-IN2P3, CENBG, F-33175 Gradignan, France}

\date{\today}% It is always \today, today,
             %  but any date may be explicitly specified

\begin{abstract}
A compact low-energy and high-intensity X-ray source for radiation biology applications is presented. A laser-induced plasma moves inside a 30 kV diode and produces a beam of 10$^{14}$ electrons at the anode location. An aluminum foil converts a part of the energy of these electrons into X-ray photons which are characterized using filtered imaging plates. The dose that would be deposited by these X-ray photons in
{\it C. elegans} larvae is calculated from Geant4 simulations. It can be set to a value ranging between 10 $\mu$Gy and 10 mGy per laser shot by simply changing the aluminum foil thickness and the diode voltage. 
Therefore, this versatile and compact X-ray source opens a new path to explore the radiation effects induced by dose rates varying over several orders of magnitude.

\end{abstract}

\maketitle

\begin{figure}
\centering
\includegraphics[height=4.cm,trim=0 0 10 0, clip=true]{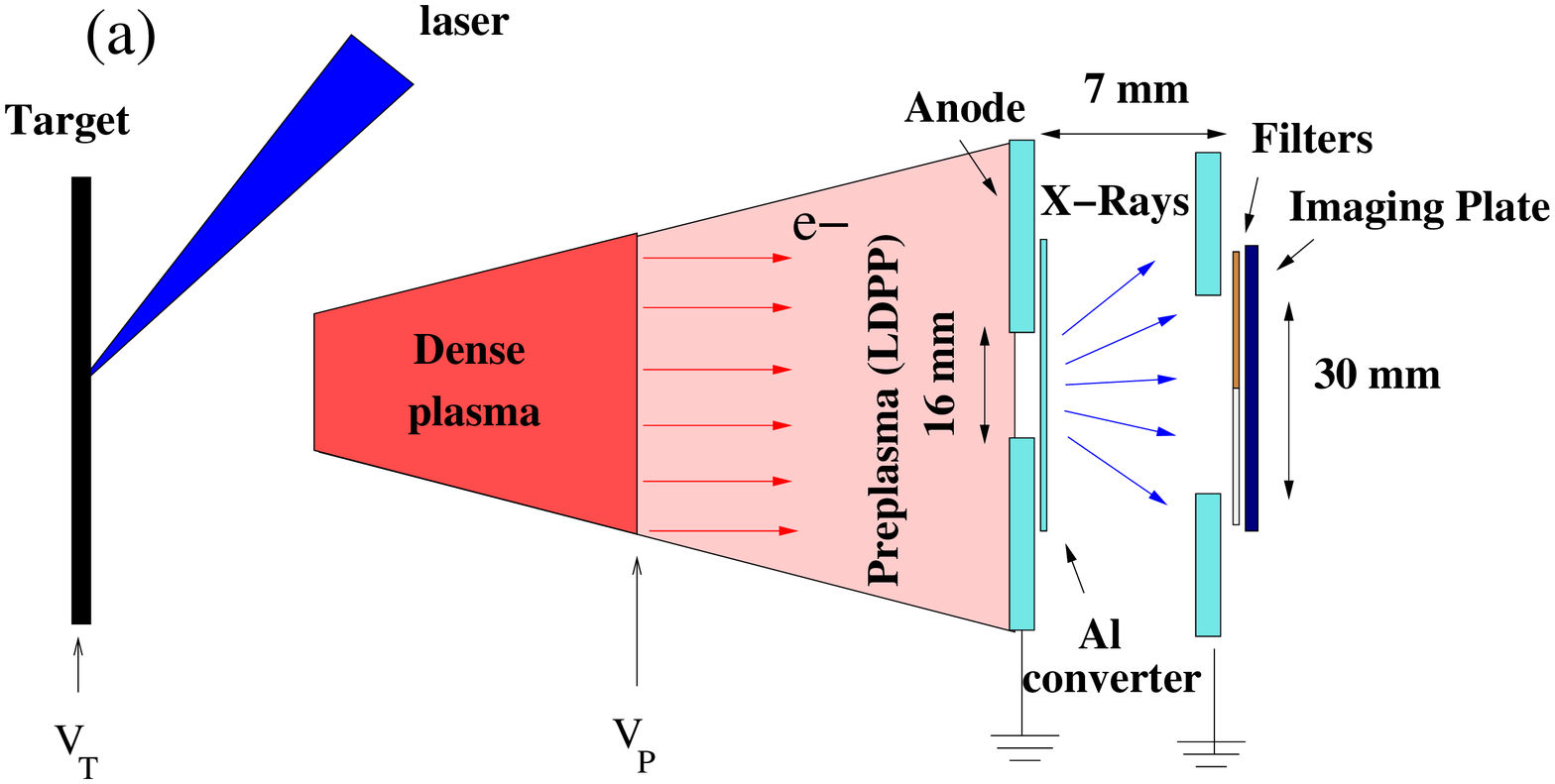}\par
\vspace{0.5cm}
\centering\includegraphics[height=5.0cm,trim=20 0 280 250, clip=true]{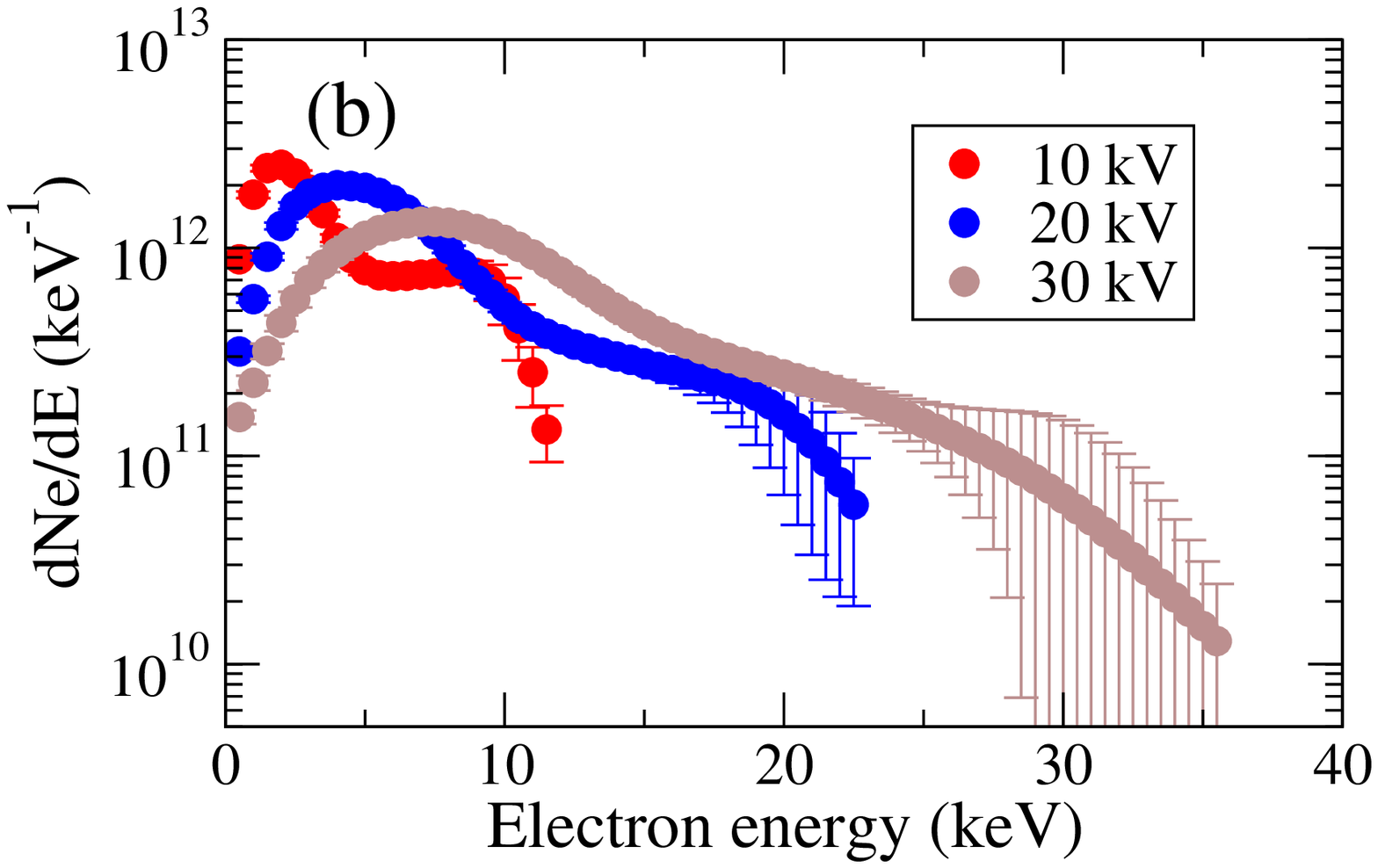}

\begin{minipage}[h]{0.40\linewidth}
\centering
\vspace{.1cm}
\includegraphics[height=3.4cm,trim=0 0 0 0, clip=true]{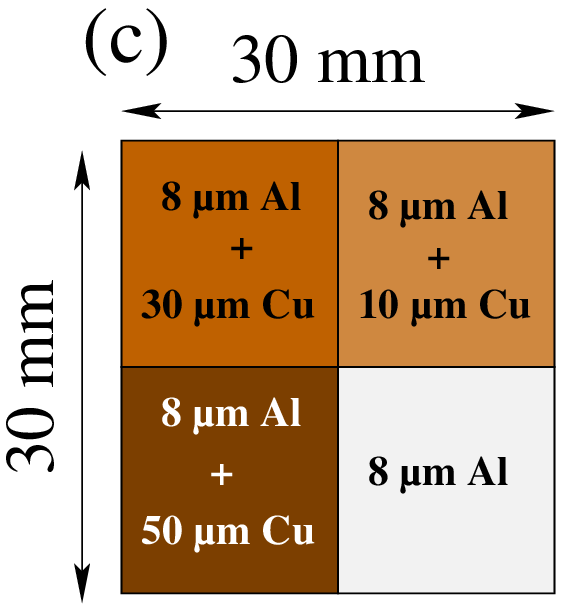}\par
\end{minipage}
\begin{minipage}[h]{0.40\linewidth}
\centering
\vspace{0.3cm}
\includegraphics[height=3.8cm,trim=0 0 0 0, clip=true]{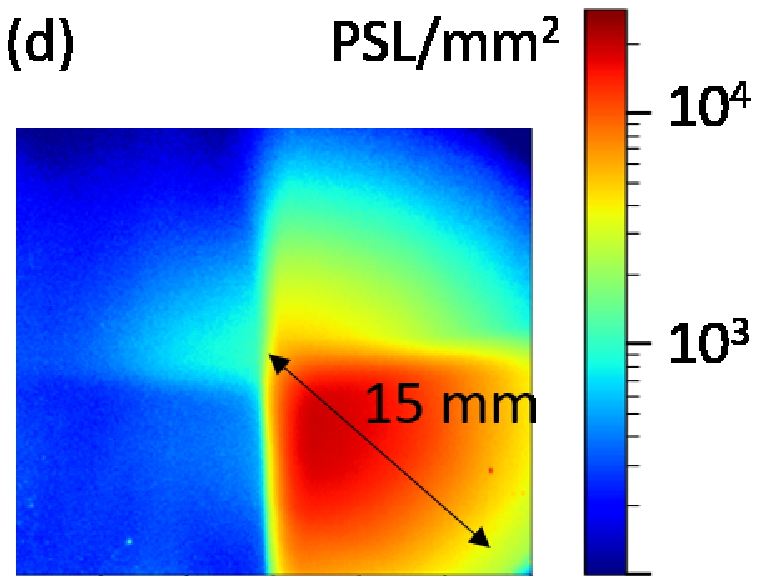}

\end{minipage}
\caption{(a) Schematic of the compact high-intensity X-ray source. The plasma positions are indicated some time after the laser shot. (b) Energy distributions of the electrons impinging the 2 cm$^2$ converter. (c) Set of filters used in front of the imaging plate (IP) to characterize X-ray photons. (d) 2-D array of photostimulated luminescence levels (PSL) of an IP exposed to X-ray photons produced in a 15 $\mu$m thick converter at V$_T$ = 25 kV.}
\end{figure}

People  are  constantly  exposed  to X-ray photons emitted from  natural terrestrial or cosmic sources\cite{katsura2016effects,pederson2020effects,vaiserman2018health,ruhm2018typical}. Dose rates of a few mGy per year are reached in these so-called chronic irradiations. Nevertheless, over a lifetime, we are subjected to acute irradiation of short duration during medical X-ray imaging or radiotherapy sessions, with dose rates exceeding several mGy per second, i.e. seven orders of magnitude higher than chronic ones.  The differences between the effects of chronic and acute irradiations on healthy cells remain one of the most controversial topics in radiation protection\cite{united2017sources,ruhm2015dose,kacem2020variation}. These studies require the development of radiation sources with wide ranges of intensity in order to perform {\it in vivo} experiments with biological models.

The nematode {\it Caenorhabditis (C.) elegans} is one of the key reference models in radiation biology\cite{sugimoto2006cell,bertucci2009microbeam,buisset2014effects,dubois2019differential,sakashita2010radiation} and cancer research\cite{kyriakakis2015caenorhabditis,kirienko2010cancer}. It has several advantages for {\it in vivo} studies: simple culture conditions and maintenance, rapid life cycle, short life span, fully sequenced genome, transparent body, only 959 somatic cells and an invariant anatomy from one animal to the next\cite{nigon2018history}. Moreover {\it C. elegans} shares numerous cellular and molecular control pathways with higher organisms, thus, biological information learned from {\it C. elegans} may be directly applicable to more complex organisms. Another interesting feature is that, in the absence of food, the L1 form of {\it C. elegans} larva can stop its development while surviving for days without molting or displaying any other morphological changes\cite{uppaluri2015size}. This starving form is interesting to study the effects of dose rate variation over a large population of identical larvae. 

The L1 stage of {\it C. elegans} has a cylindrical shape with a length of 250 $\mu$m and a radius of 6 $\mu$m.The primary endpoint during nematode X-ray irradiation concerns the effects on reproduction that occur for dose rates ranging from 10 to 50 mGy/h\cite{maremonti2019gamma}. Considering X-ray photons of 10 keV, a simple calculation shows that a dose of the order of 1 mGy can be reached with approximately 5000 photons incident onto the larva body. In this context, high-fluence X-ray sources are necessary to irradiate a large quantity of {\it C. elegans}. Studies at dose rates lower than 10 mGy/s are usually performed with standard radioactive sources such as $^{137}$Cs. Considering the number of required X-ray photons, strong activities of the order of 1 TBq are needed leading to severe radio-protection issues.

Low-energy and high intensity X-ray sources can also be produced either in direct laser-plasma interaction\cite{li2017laser} or with plasma focus devices\cite{sumini2015analysis}. With X-ray fluxes up to several 10$^{10}$ photons.sr$^{-1}$.s$^{-1}$ and small spot area of 100 µm$^2$, X-ray sources based on direct laser-plasma interactions open new perspectives in medical imaging \cite{li2017laser}. A similar number of X-ray photons can be emitted from a large-spot plasma focus devices in a single shot. They can deliver dose rates of several Gy in a few tens of ns in cells, allowing applications in radiotherapy \cite{sumini2015analysis}. However, the dose rate produced by all these devices cannot be easily adjusted over a wide range, thereby limiting \textit{in vivo} experiments with biological models. By biasing a laser-induced plasma, we show in this letter that it is possible to generate a compact source of low-energy, high-intensity X-ray photons with the required versatility of biology applications.

We have recently developed a 10 Hz electron plasma source capable of delivering 10$^{14}$ electrons per bunch with kinetic energy up to 30 keV\cite{gobet2020versatile}. In this Letter we describe an application of this electron device as a compact X-ray source. We have characterized the latter and we show that up to 10$^9$ photons are produced per electron bunch. The amount of X-ray photons can be controlled and doses can be delivered in a few mm$^3$ solution of L1 {\it C. elegans} larvae, ranging from 10 $\mu$Gy to 10 mGy, this in single shot operation of the electron source. 

The main characteristics of the electron source were already carefully described\cite{gobet2020versatile,raymond2017energy,verst,comet2016absolute} and are briefly summarized in the following. A schematic of the device is displayed in Fig.1(a). A 10 ns, 10$^{13}$ W/cm$^{2}$ Nd:YAG laser pulse is focused on an aluminum target at a repetition rate up to 10 Hz. Each shot produces a plasma in which about 2x10$^{15}$ electrons can be released. This plasma presents two components: a dense aluminum plasma, with density reaching 10$^{20}$-10$^{22}$ part.m$^{-3}$, preceded by a Low Density anisotropic Pre-Plasma (LDPP), containing approximately 10$^{16}$-10$^{17}$ part.m$^{-3}$. This two-components plasma expands during 130 ns between the aluminum  target, biased at a negative voltage -$V_T$, and the 2 mm thick anode plate located 50 mm downstream from the target. 

Electrons are extracted and accelerated from the front end boundary of the dense plasma component. In Fig.1(b), we have reported,  for 3 target voltages, the measured energy distributions of the electrons that have reached the 2 cm$^2$ central area of the anode. The number of incident electrons increases from 10$^{13}$ at V$_T$=10 kV to 1.6 10$^{13}$ at V$_T$=30 kV. These distributions are continuous and indicate a maximum energy greater than eV$_T$. Indeed, the first extracted electrons are accelerated toward the anode by the electric field induced by V$_T$ but they are also pushed by their followers, allowing them to gain additional kinetic energy. 

The electrons produce X-ray photons as follows. A circular 16 mm-diameter hole is drilled in the central part of the anode plate and a grounded thin aluminum secondary target , 15 or 100 µm thick, is placed just behind this hole. The thickness is large enough to stop all electrons by collisions (atomic excitation or ionization) or by braking radiation (Bremsstrahlung process). The efficiency of the latter process is approximately 10$^{-4}$ for 10 keV electrons\cite{estar}. Because of multiple scattering of the electrons during their slowing down, the X-ray emission is isotropic and lasts as the electron bunch ($\sim$ 30 ns FWHM). Moreover, these secondary aluminum targets, called converters in the following are thin enough to allow most of the forward emitted radiation to exit the foil.

In the reported experiment, the X-ray photons are detected by using Fujifilm$^{TM}$ BAS-SR imaging plates (IP) located 7 mm behind the converter. IPs store incident X-ray energy in phosphor elements that are directly read into units of PhotoStimulated Luminescence level (PSL)\cite{von1992x} by using a dedicated scanner. Spectral sensitivity of IPs is broad\cite{bonnet2013response2} and they respond linearly to X-ray fluence\cite{zeil2010absolute} with a high dynamic range. In addition, they are insensitive to electromagnetic noise. In this experiment,  the IP covers a circular area of radius 15 mm and is topped with a square filter, as shown in Fig.1(c). The filter is divided in four aluminum/copper sheets of different thicknesses:  8 $\mu$m Al, 8 $\mu$m Al + 10 $\mu$m Cu, 8 $\mu$m Al + 30 $\mu$m Cu and 8 $\mu$m Al + 50 $\mu$m Cu. They are at least 50\% transparent to photons above the following energies: 4, 15, 22 and 26 keV, respectively.

After exposure to one X-ray shot, the IP is removed from the experimental apparatus. The scan is performed two minutes after the laser shot and a correction to spontaneous decay (fading process) is applied taking into account this delay\cite{bonnet2013response}. Each scan produces a 2-D array of PSL values. A typical image of an irradiated IP at 25 kV is displayed in Fig.1(d): the redder the color, the higher the PSL surface density. As expected, the PSL density decreases with filter thickness. The 15 mm-radius circular structure corresponding to the area of the IP exposed to X-ray photons is observed. The scanned images of the IPs are shown in Fig.2(a) (top) at different source voltages. The X-ray energy deposition, as observed by the PSL, increases with the target voltage because the energy as well as the number of incident electrons scale with the 
latter\cite{gobet2020versatile}.

\begin{figure}
\centering
\includegraphics[height=4.5cm,trim=0 0 0 0, clip=true]{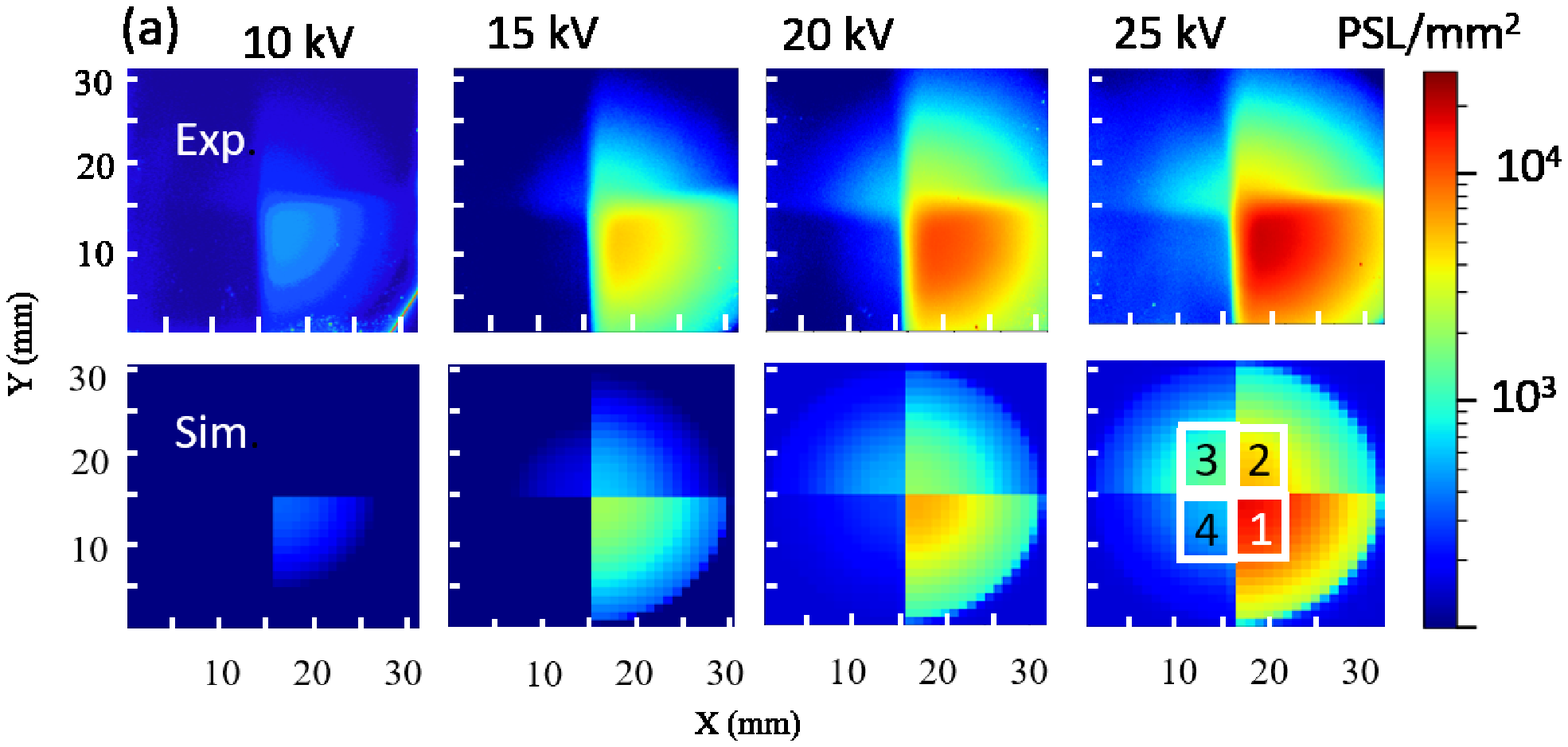}
\begin{minipage}[h]{0.47\linewidth}
\includegraphics[height=5.cm,trim=20 100 450 80, clip=true]{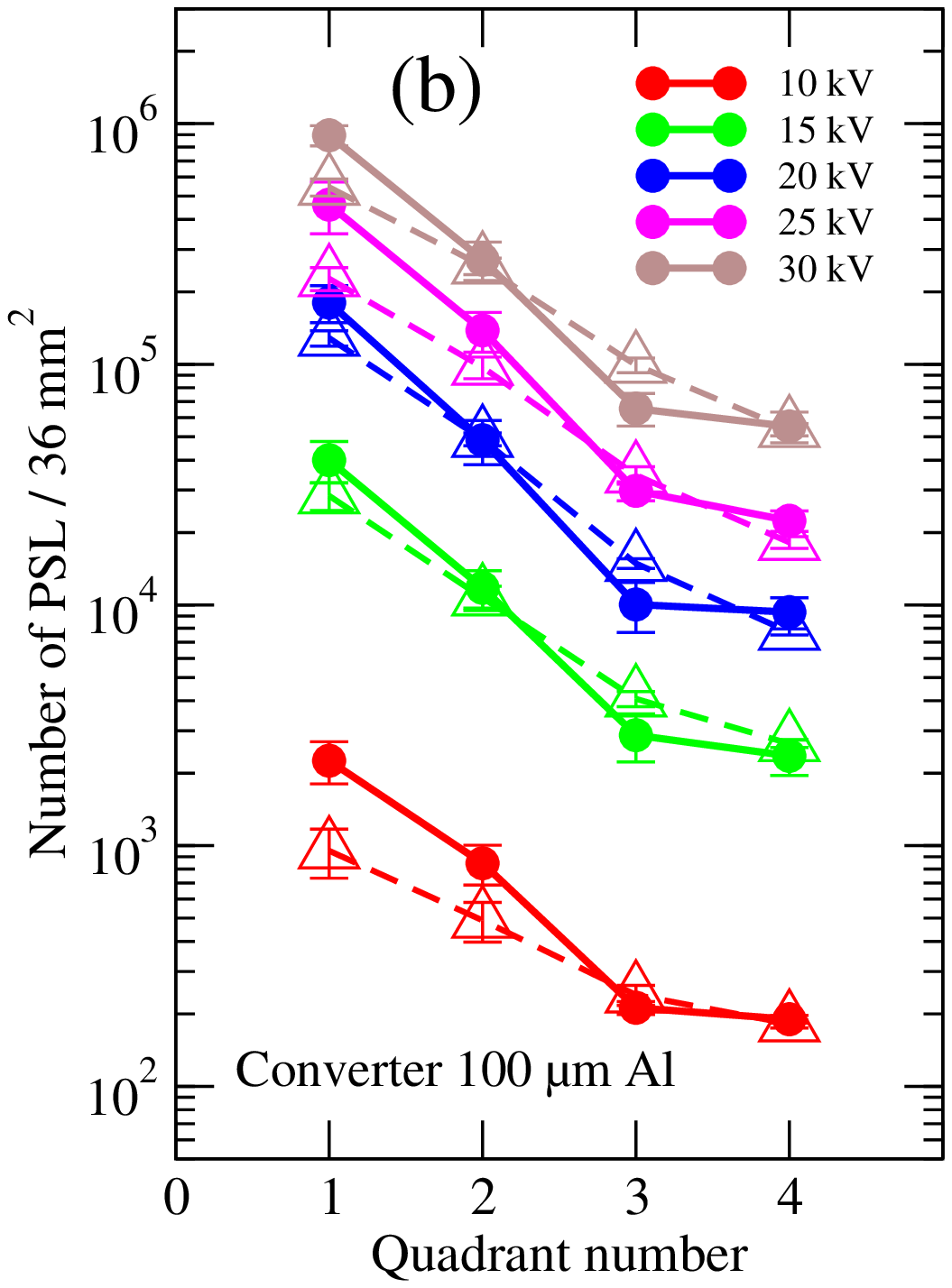}
\end{minipage}
\begin{minipage}[h]{0.47\linewidth}
\includegraphics[height=5.cm,trim=20 100 450 80, clip=true]{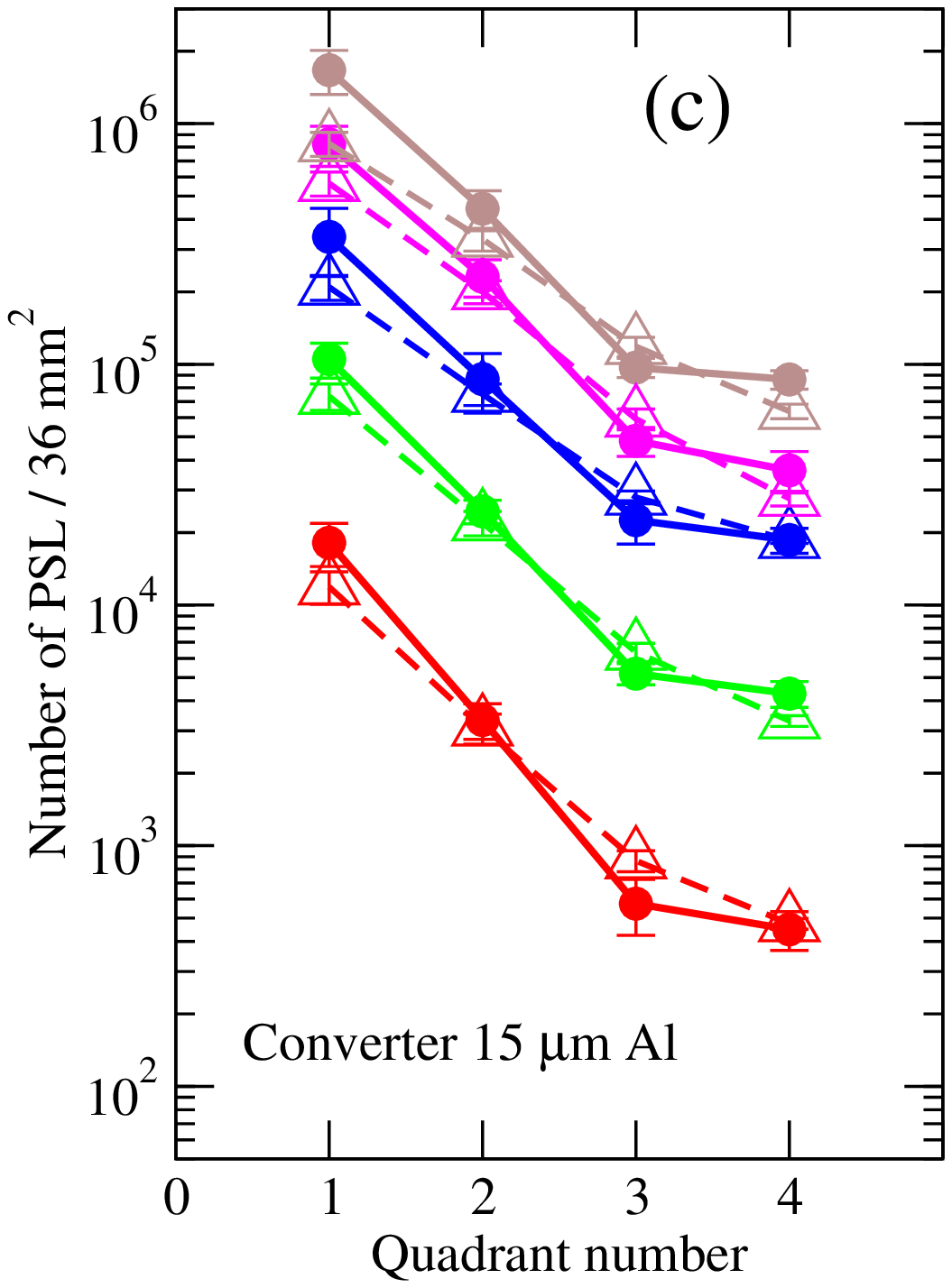}
\end{minipage}
\caption{(a) 2-D array of PSL values of an IP exposed to X-ray photons produced at various values of V$_T$ with a 15 $\mu$m thick converter, top: experiment, bottom: simulation. (b,c) Integrals of the number of PSL in the four square quadrants defined in Fig.2a (bottom right) for an aluminum converter (b) 100 $\mu$m thick  or (c) 15 $\mu$m thick, solid symbols: experiment, hollow symbols: simulation.}
\end{figure}

We reproduced these scans by performing Geant4 Monte Carlo simulations. We have used its 10.4.p01 version \cite{agostinelli2003geant4} associated with the Livermore Physics List \cite{allison2016recent}. This allowed us to compute particle interactions (electrons, photons) with matter at energies as low as 100 eV.  For electrons, ionization, Bremsstrahlung and multiple scattering processes are considered. For photons, the photoelectric effect and Compton scattering are taken into account. The geometry of the  simulation includes the Al converter, the filters and the IP internal structure\cite{bonnet2013response,bonnet2013response2}. All material characteristics come from the National Institute Standards and Technology material lists. The energy distributions of electrons reported in Fig.1(b) are used as input. One billion simulated electrons interacting with the al converter is a reasonable trade-off to save calculation time while maintaining statistical error less than 1\%. The electron beam impinges the aluminum converter perpendicular to surface and produces X-ray photons. For each photon reaching the IP, the energy deposited in the sensitive layer is stored, along with its spatial position. Then, the energy deposition profile is used along with the known IP response function\cite{bonnet2013response2}. Finally, a normalization factor has been applied to the results to consider the number of incident electrons onto the Al converter as measured experimentally. We present the simulated 2-D arrays of PSL values in Fig.2(a) (bottom). In order to compare with the actual images displayed in Fig.2(a) (top), we have added the experimental background contribution estimated from the non-exposed part of the IPs (radii>15 mm). A good agreement, both in structure and PSL density, is observed.
 
For each 2-D array, we have summed the number of PSLs in the four 36 mm$^2$ square quadrants defined in Fig.2(a) (bottom right). The mean values over 3 shots, along with their standard deviation, are given in Fig.2(b) and Fig.2(c) both for the experimental images (solid symbols) and for the calculated maps (hollow symbols). The PSL yield varied within approximately 3 orders of magnitude by changing the converter thickness or the target voltage. Geant4 simulations
agree very well with the observations. We have plotted in Fig.3 the predicted X-ray energy distributions inside the 706 mm$^2$ central area of the IP for the two 15 $\mu$m and 100 $\mu$m thick converters and at the previously mentioned voltages. The spectra are continuous above 2 keV, confirming the large tunability of Bremsstrahlung X-ray production. 
The low-energy X-ray photons at $\sim$ 1.5 keV are K$_\alpha$ radiations emitted from aluminum atoms excited by incident electrons. The X-ray photons are mostly produced in the first 10 µm of the converter and have a great probability to be absorbed in thick targets. Consequently, the average energy of the X-ray photons produced in the 100 $\mu$m thick Al converter is higher than for the 15 $\mu$m thick one. In summary, the X-ray source described in this paper is capable of producing a large amount of photons per electron bunch: from a few 10$^6$ in the 100 $\mu$m thick converter at V$_T$ = 10 kV up to a few 10$^9$ photons in the 15 $\mu$m converter at V$_T$ = 30 kV. The spot area of this source is 200 mm$^2$. In a first approximation, the amount of X-ray photons depends linearly on the area of anode hole.

\begin{figure}
\centering
\includegraphics[height=8cm,trim=0 40 0 0, clip=true]{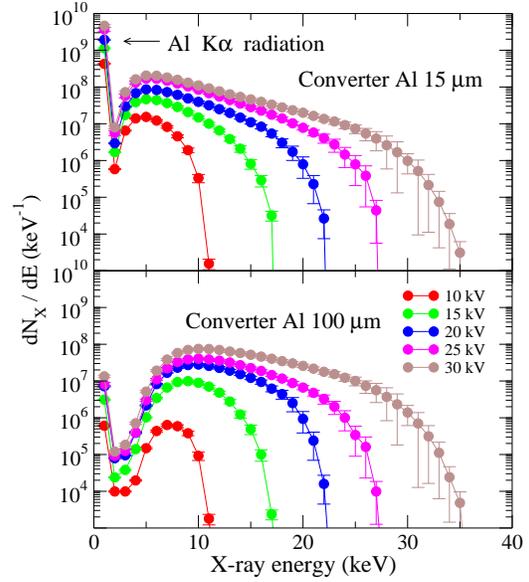}
\caption{Calculated energy distributions of X-ray photons impinging the IP at different target voltages and for two converter thicknesses.}
\end{figure}

\begin{figure}
\centering
\includegraphics[height=3.5cm,trim=0 0 0 0, clip=true]{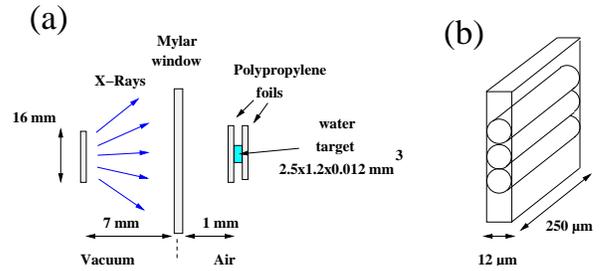}\par
%\begin{minipage}[h]{0.47\linewidth}
%\includegraphics[height=3.5cm,trim=0 0 0 0, clip=true]{set_up_geant4_celegans_irradiation.eps}\par
%\end{minipage}
%\begin{minipage}[h]{0.47\linewidth}
%\includegraphics[height=3.5cm,trim=0 0 0 0, %clip=true]{zoom_celegans_model.eps}\par
%\end{minipage}

\caption{(a) Geometrical configuration used in the Geant4 simulation for radiobiological applications. (b) Stack of C-Elegans models in the water target.}
\end{figure}

Let us now close the under-vacuum X-ray source by a 50 $\mu$m thick Mylar foil window. Then, install in air, a few millimeters downstream from this window, the sample to be irradiated. We then obtain the simplest configuration of an X-ray irradiation facility for radiobiology applications as mentioned in the beginning of this letter. In a Geant4 simulation we have considered a 0.036 mm$^3$ droplet of water solution, containing a population of 1000 larvae of C.{\it elegans},  embedded between two 10 $\mu$m thick sheets of polypropylene and installed in air 1 mm  behind the Mylar window as presented in Fig.4(a). The nematodes are modeled by stacks of 12 $\mu$m diameter, 250 $\mu$m long water cylinders as shown in Fig.4(b). The simulated X-rays are emitted isotropically from the converter position with the energy distributions presented in Fig.3. We compute the energy deposited  in each cylinder by the photons emitted in one shot of the X-ray source. Then, by dividing this energy by the mass of each cylinder, we determine the dose deposited in the population of larvae. The results are displayed in Fig.5 in logarithmic scale for the two converters and at the usual target voltages. The doses are computed for an increasing number of X-ray shots ranging from 5 to 160 in a geometric progression of factors of 2. The maximum value of each distribution is normalized to unity in order to facilitate the analysis of the curves. We observe the relative width of the distributions decreases with the number of shots. Actually, the number of X-ray photons that deposit their energy in a nematode increases, thus reducing the statistical dispersion. For 160 laser shots, the energy deposition dispersion is only a few percent, regardless converter thickness or target voltage. In addition, the maxima of the distributions are shifted by a constant factor each time the number of shots is doubled. That means the average dose deposited in a larvae population would increase linearly with the number of shots.

\begin{figure}
\centering
\includegraphics[height=7.cm,trim=0 70 200 10, clip=true]{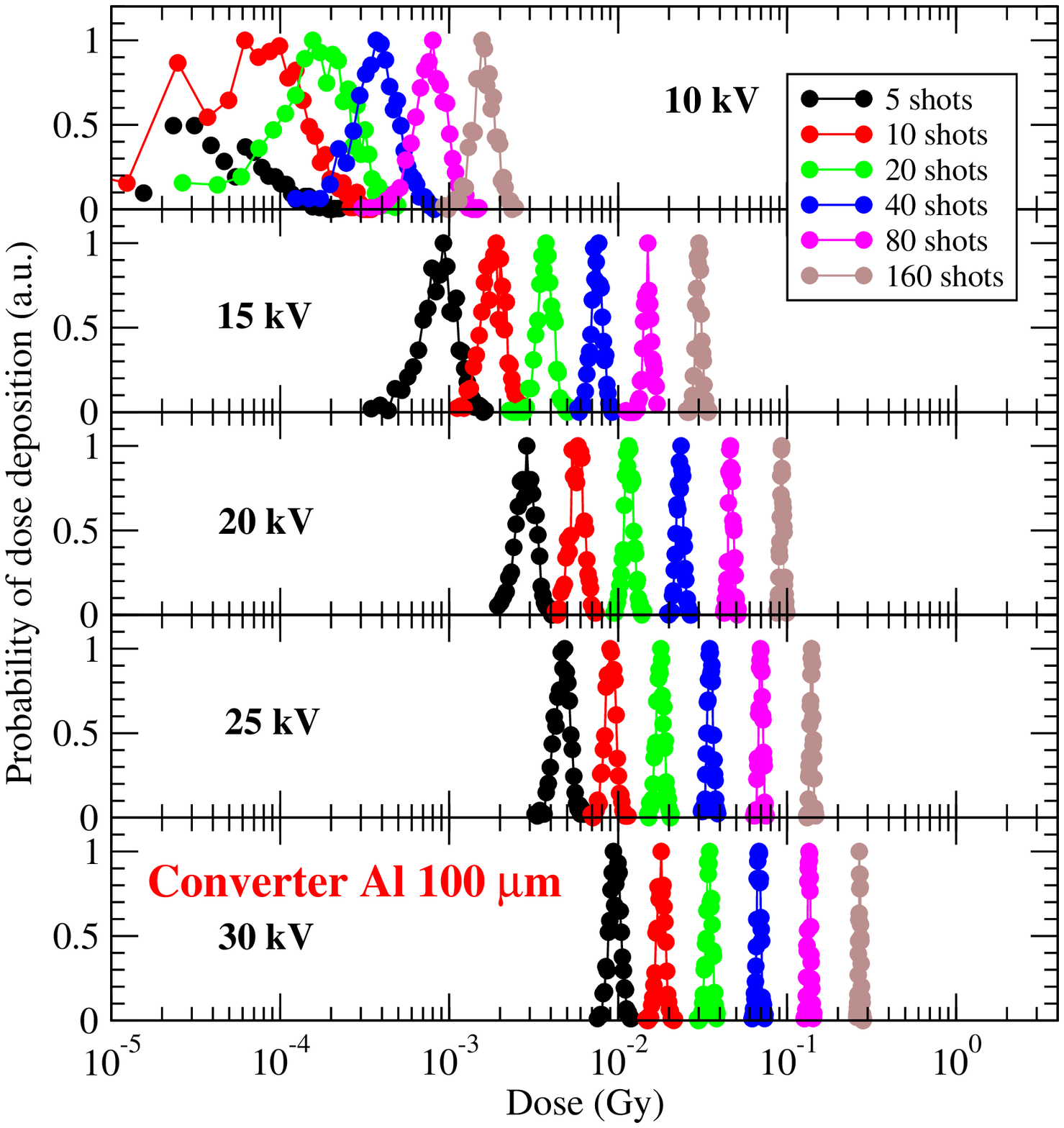}
\includegraphics[height=7.cm,trim=0 70 200 10, clip=true]{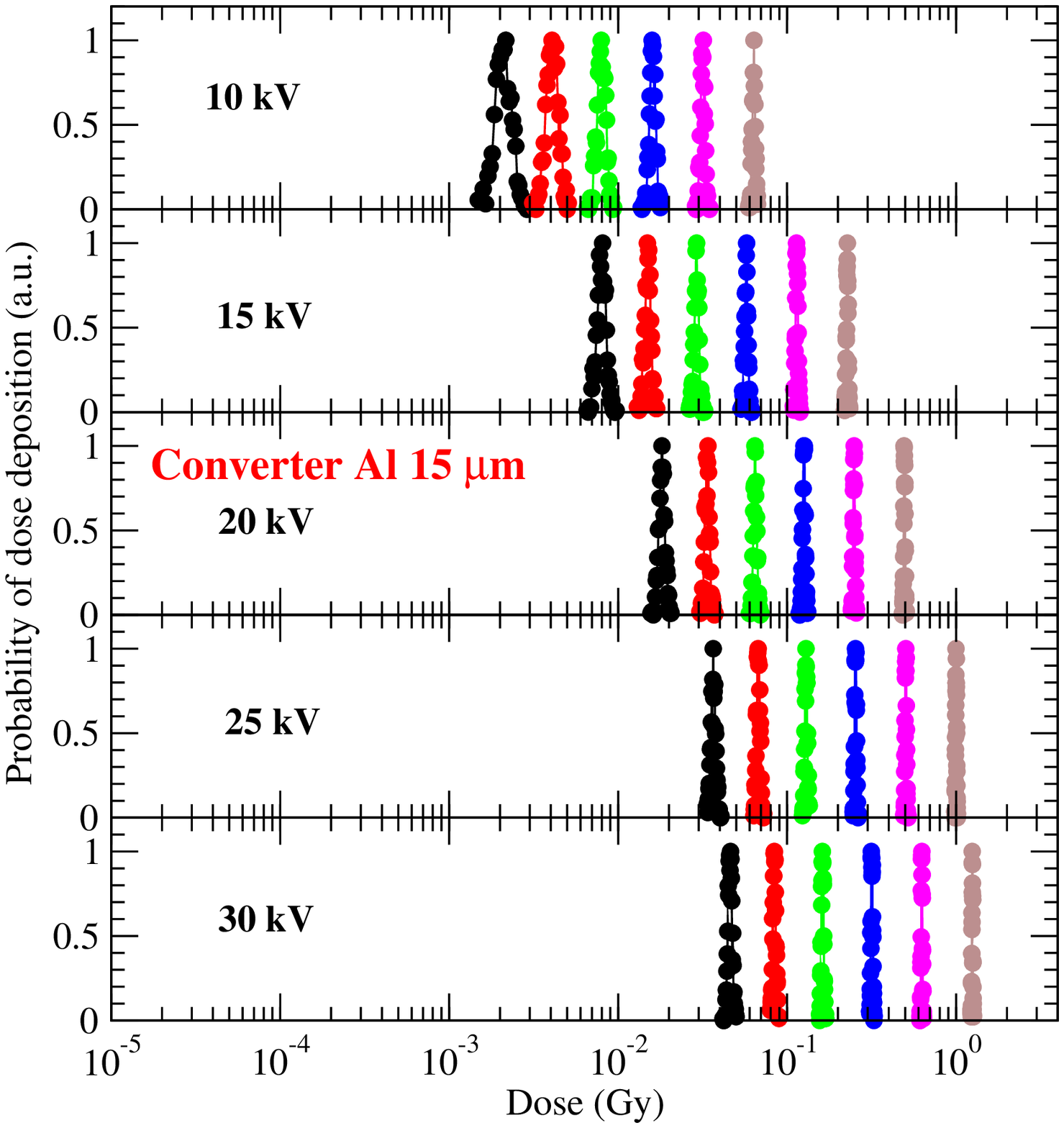}
\caption{Simulated distributions of dose deposition in C-elegans nematodes at different target voltages and for two converter thicknesses.}
\end{figure}

\begin{figure}
\centering
\includegraphics[height=5.5cm,trim=20 50 250 180, clip=true]{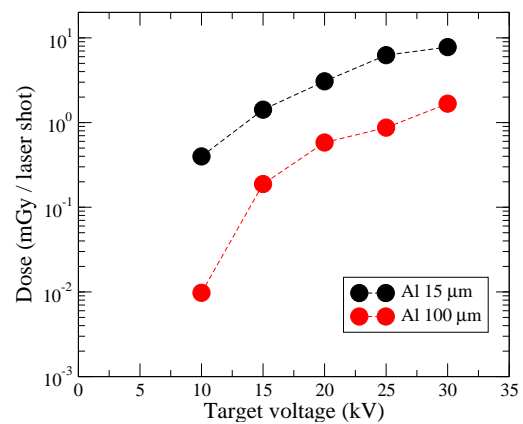}
\caption{Calculated dose deposited in the nematode per laser shot at different target voltages and for two converter thicknesses.}
\end{figure}

The average dose accumulated in the larva of nematode per laser shot is displayed in Fig.6. It has been deduced from the dose reached after 160 shots by a simple division and the uncertainty on the calculated value is smaller than the symbol size. It increases monotonically with the target voltage over almost 3 orders of magnitude for the two converters considered in this study. It reaches 10 $\mu$Gy per shot for a 100 $\mu$m thick aluminum converter at a voltage of 10 kV, up to 10 mGy per shot for a 15 $\mu$m converter at a voltage of 30 kV. Additional numerical Geant4 simulations were performed with a thicker water target (100 $\mu$m instead of 12 $\mu$m). This size is closer to the actual experimental dimensions in biological experimental tests. If the C.{\it elegans} larvae are installed on the rear part of the new target, they receive a reduced dose due to photon absorption in water. The results are nevertheless similar to the previous ones with a decrease by a factor 2 of the dose accumulated in C.{\it elegans} per laser shot (approximately the size of the symbol used in Fig.6).

In summary, we have analyzed in this letter the main characteristics of the X-ray photons that could be produced in the already published compact laser-induced plasma electron gun. The X-ray yield spans over three orders of magnitude and the dose available per laser shot in a nematode cell can be set in a range from 10 $\mu$Gy up to 10 mGy. With a 100 $\mu$m thick aluminum converter, a target voltage of 10 kV, and a repetition rate of 1 Hz, a dose rate of the order of 30 mGy/h can be reached allowing for accurate studies of the most sensitive endpoints in radiobiological studies such as reproduction\cite{maremonti2019gamma}. Furthermore, we could improve the present device by using a thinner converter to reduce the X-ray self-attenuation. In this case, at 30 kV and with a laser repetition rate of 10 Hz, radiobiological effects at dose rates of up to 100 mGy/s can be addressed without the radioprotection issues related to the high activity of standard radioactive sources. It is easy to change the repetition rate, the converter thickness or the target voltage, making this compact X-ray source versatile and potentially very useful to study radiobiological effects. In particular, the effects of low-to-high dose rate irradiations over more than 5 orders of magnitude could be addressed. This X-ray source could also be of prime interest for other applications or studies such as energy relaxation in scintillators\cite{tantot2013sound}. We can mention the unexpected long-lived light emission (afterglow) recently observed in LaBr3  crystals\cite{tarisien2018scintillators} under X and $\gamma$-ray flashes that could be investigated in detail thanks to a controlled amount of deposited X-ray energy ranging from 1 nJ to 1 $\mu$J. 

\begin{acknowledgments}
The authors acknowledge Dr. D. Smith for careful reading of the manuscript.
\end{acknowledgments}

\medskip

\textbf{DATA AVAILABILITY}

\medskip

The data that support the findings of this study are available from the corresponding author upon reasonable request.

\bibliographystyle{apsrev4-1}

\bibliography{biblio}

\end{document}